%% file: DirectTestArrowDebreuMathFi.tex
\documentclass[12pt]{article}
\usepackage{fullpage,times,graphicx,amssymb,psfrag,setspace}

\input defs.tex

\usepackage[agsmcite]{harvard}
\renewcommand{\cite}{\citeasnoun}
\onehalfspace

\begin{document}

\title{\textsc{\large{A market test for the positivity of Arrow-Debreu prices}}}
\author{By Alexandre d'Aspremont\footnote{
ORFE, Princeton University, Princeton, NJ 08544, USA. alexandre.daspremont@m4x.org.
\iffalse This work started as a part of the author's thesis at Stanford University under the direction of Stephen Boyd. The author would also like to thank Patrick Cheridito, Rama Cont, Darrell Duffie, Hans F\"ollmer, Laurent El Ghaoui, Paul Glasserman, David Heath, Nicole El Karoui, Noureddine El Karoui as well as participants to the NSF sponsored Intitute for Mathematics and its Applications workshop on Risk Management and Model Specifications Issues in Finance, the Petits D\'ejeuners de la Finance in Paris, the Stochastic Analysis Seminar at Princeton University and the INFORMS 2005 Applied Probability Conference in Ottawa. XXX acknowledge funding XXX \fi}} 
\maketitle

\begin{abstract}
We derive tractable necessary and sufficient conditions for the absence of static or buy-and-hold arbitrage opportunities in a perfectly liquid, one period market. We formulate the positivity of Arrow-Debreu prices as a generalized moment problem to show that this no arbitrage condition is equivalent to the positive semidefiniteness of matrices formed by the market prices of tradeable securities and their products. We apply this result to a market with multiple assets and basket call options.
\end{abstract}

\vskip 1ex
\textbf{Keywords: }Arbitrage, moment conditions, basket options, semidefinite programming, harmonic analysis on semigroups.

\section{Introduction}
The fundamental theorem of asset pricing establishes the equivalence between absence of arbitrage and existence of a martingale pricing measure, and is the foundation of the \cite{Blac73} and \cite{Mert73} option pricing methodology. Option prices are computed by an arbitrage argument, as the value today of a dynamic, self-financing hedging portfolio that replicates the option payoff at maturity. This pricing technique relies on at least two fundamental assumptions: it posits a model for the asset dynamics and assumes that markets are frictionless, i.e. that continuous trading in securities is possible at no cost. Here we take the complementary approach: we do not make any assumption on the asset dynamics and we only allow trading today and at maturity. In that sense, we revisit the classic result of \cite{Arro54} on the equivalence between positivity of state prices and absence of arbitrage in a one period market. In this simple market, we seek computationally \emph{tractable} conditions for the absence of arbitrage, directly formulated in terms of \emph{tradeable} securities.

Of course, these results are not intended to be used as a pricing framework in liquid markets. Our objective here instead is twofold. First, market data on derivative prices, aggregated from a very diverse set of sources, is always plagued by liquidity and synchronicity issues. Because these price data sets are used by derivatives dealers to calibrate their models, we seek a set of arbitrarily refined tests to detect unviable prices in the one period market or, in other words, detect prices which would be incompatible with \emph{any} arbitrage free dynamic model for asset dynamics. Second, in some very illiquid markets, these conditions form simple upper or lower hedging portfolios and diversification strategies that are, by construction, immune to model misspecification and illiquidity issues. 

Work on this topic starts with the \cite{Arro54} no arbitrage conditions on state prices. This was followed by a stream of works on multiperiod and continuous time extensions stating the equivalence between existence of a martingale measure and absence of dynamic arbitrage, starting with \cite{Harr79} and \cite{Harr81}, with the final word probably belonging to \cite{Dala90} and \cite{Delb05}. Efforts to express these conditions directly in terms of asset prices can be traced back to \cite{Bree78} and \cite{Frie79} who derive equivalent conditions on a continuum of (possibly nontradeable) call options. \cite{Bree78}, \cite{Jack96} and \cite{Laur00} use these results to infer information on the asset distribution from the market price of calls using a minimum entropy approach. A recent paper by \cite{Davi05} provides explicit no arbitrage conditions and option price bounds in the case where only a few single asset call prices are quoted in a multiperiod market. Finally, contrary to our intuition on static arbitrage bounds, recent works by \cite{Hobs04} and \cite{dasp02c} show that these price bounds are often\emph{very close} to the price bounds obtained using a Black-Scholes model, especially so for options that are outside of the money.

Given the market price of tradeable securities in a one period market, we interpret the question of testing for the existence of a state price measure as a generalized moment problem. In that sense, the conditions we obtain can be seen as a direct generalization of Bochner-Bernstein type theorems on the Fourier transform of positive measures. Market completeness is then naturally formulated in terms of moment determinacy. This allows us to derive equivalent conditions for the absence of arbitrage between \emph{general payoffs} (not limited to single asset call options). We also focus on the particular case of basket calls or European call options on a basket of assets. Basket calls appear in equity markets as index options and in interest rate derivatives market as spread options or swaptions, and are key recipients of market information on correlation.

The paper is organized as follows. We begin by describing the one period market and illustrate our approach on a simple example, introducing the payoff semigroup formed by the market securities and their products. Section 2 starts with a brief primer on harmonic analysis on semigroups after which we describe the general no arbitrage conditions on the payoff semigroup. We also show how the products in this semigroup complete the market. We finish in Section 3 by a case study on spread options.

\subsection{One Period Model}
We work in a one period model where the market is composed of $n$ assets with payoffs at maturity equal to $x_i$ and price today given by $p_i$ for $i=1,\ldots,n$. There are also $m$ derivative securities with payoffs $s_j(x)=s_j(x_1,\ldots,x_n)$ and price today equal to $p_{n+j}$ for $j=1,\ldots,m$. Finally, there is a riskless asset with payoff 1 at maturity and price 1 today and we assume, without loss of generality here, that interest rates are equal to zero (we work in the forward market).
We look for conditions on $p$ precluding arbitrage in this market, i.e. buy and hold portfolios formed at no cost today which guarantee a strictly positive payoff at maturity. 

We want to answer the following simple question: Given the market price vector $p$, is there an arbitrage opportunity (a buy-and-hold arbitrage in the continuous market terminology) between the assets $x_i$ and the securities $s_j(x)$?
Naturally, we know from the \cite{Arro54} conditions that this is equivalent to the existence of a state price (or probability) measure $\mu$ with support in $\reals^n_+$ such that:
\BEQ
\label{eq:exist-proba}
\BA{l}
\Expect_\mu[x_i]=p_i,\quad i=1,\ldots,n,\\
\Expect_\mu[s_j(x)]=p_{n+j},\quad j=1,\ldots,m,\\
\EA
\EEQ
\cite{Bert00} show that this simple, fundamental problem is computationally hard (in fact NP-Hard). In fact, if we simply discretize the problem on a uniform grid with $L$ steps along each axis, this problem is still equivalent to an exponentially large linear program of size $O(L^n)$. Here, we look for a discretization that does not involve the state price measure but instead formulates the no arbitrage conditions directly on the market price vector $p$. Of course, NP-Hardness means that we cannot reasonably hope to provide an efficient, exact solution to all instances of problem (\ref{eq:exist-proba}). Here instead, we seek an arbitrarily refined, computationally efficient relaxation for this problem and NP-Hardness means that we will have to tradeoff precision for complexity. 

\subsection{The Payoff Semigroup}
To illustrate our approach, let us begin here with a simplified case were $n=1$, i.e. there is only one forward contract with price $p_1$, and the derivative payoffs $s_j(x)$ are monomials with
$s_j(x)=x^{j}$ for $j=2,\ldots,m$. In this case, conditions (\ref{eq:exist-proba}) on the state price measure $\mu$ are written:
\BEQ
\label{eq:exist-proba-stieljes}
\BA{l}
\Expect_\mu[x^{j}]=p_j,\quad j=2,\ldots,m,\\
\Expect_\mu[x]=p_1,\\
\EA
\EEQ
with the implicit constraint that the support of $\mu$ be included in $\reals_+$. We recognize (\ref{eq:exist-proba-stieljes}) as a Stieltjes moment problem. For $x\in\reals_+$, let us form the column vector $v_m(x)\in\reals^{m+1}$ as follows:
\[
v_m(x)\triangleq(1,x,x^2,\ldots,x^{m})^T.
\]
For each value of $x$, the matrix $P_m(x)$ formed by the outer product of the vector $v_m(x)$ with itself is given by:
\[
P_m(x)\triangleq v_m(x)v_m(x)^T=
\left(\BA{cccc}
1 & x & \ldots & x^{m}\\
x & x^2 &  & x^{m+1}\\
\vdots &  & \ddots & \vdots \\
x^{m} & x^{m+1} & \ldots & x^{2m}\\
\EA\right)
\]
$P_m(x)$ is a \emph{positive semidefinite} matrix (it has only one nonzero eigenvalue equal to $\|v_m(x)\|^2$). If there is no arbitrage and there exists a state price measure $\mu$ satisfying the price constraints (\ref{eq:exist-proba-stieljes}), then there must be a symmetric moment matrix $M_m\in\reals^{(m+1)\times(m+1)}$ such that:
\[
M_m\triangleq \Expect_\mu[P_m(x)]=
\left(\BA{cccc}
1 & p_1 & \ldots & p_{m}\\
p_1 & p_2 &  & \Expect_\mu[x^{m+1}]\\
\vdots &  & \ddots & \vdots\\
p_{m} & \Expect_\mu[x^{m+1}] & \ldots & \Expect_\mu[x^{2m}]\\
\EA\right)
\]
and, as an average of positive semidefinite matrices, $M_m$ must be positive semidefinite. In other words, the existence of a positive semidefinite matrix $M_m$ whose first row and columns are given by the vector $p$ is a necessary condition for the absence of arbitrage in the one period market. In fact, positivity conditions of this type are also \emph{sufficient} (see \cite{Vasi02} among others). Testing for the absence of arbitrage is then equivalent to solving a \emph{linear matrix inequality}, i.e. finding matrix coefficients corresponding to $\Expect_\mu[x^{j}]$ for $j=m+1,\ldots,2m$ that make the matrix $M_m(x)$ positive semidefinite.

This paper's central result is to show that this type of reasoning is not limited to the unidimensional case where the payoffs $s_j(x)$ are monomials but extends to arbitrary payoffs. Instead of looking only at monomials, we will consider the \emph{payoff semigroup} $\mathbb{S}$ generated by the payoffs 1, $x_i$ and $s_j(x)$ for $i=1,\ldots,n$ and $j=1,\ldots,m$ and their products (in graded lexicographic order):
\BEQ
\label{eq:semigroup}
\mathbb{S}\triangleq\left\{1,x_1,\ldots,x_n,s_1(x),\ldots,s_m(x),x_1^2,\ldots,x_is_j(x),\ldots,s_m(x)^2,\ldots\right\}
\EEQ
In the next section, we will show that the no arbitrage conditions (\ref{eq:exist-proba}) are equivalent to positivity conditions on matrices formed by the prices of the assets in $\mathbb{S}$. We also detail under which technical conditions the securities in $\mathbb{S}$ make the one period market complete. In all the results that follow, we will assume that the asset distribution has \emph{compact support}. As this can be made arbitrarily large, we do not loose much generality from a numerical point of view and this compactness assumption greatly simplifies the analysis while capturing the key link between moment conditions and arbitrage. Very similar but much more technical results hold in the non compact case, as detailed in the preprint \cite{dasp03c}.

\subsection{Semidefinite Programming}
\label{ss:sdp} The key incentive for writing the no arbitrage conditions in terms of linear matrix inequalities is that the later are \emph{tractable}. The problem of finding coefficients that make a particular matrix positive semidefinite can be written as:
\BEQ
\label{eq:sdp}
\BA{ll}
\mbox{find} & {y}\\
\mbox{such that} & C + \sum_{k=1}^m y_k A_k \succeq 0\\
\EA
\EEQ
in the variable $y\in\reals^m$, with parameters $C,~A_k\in\reals^{n\times n}$, for $k=1,\ldots,m$, where $X\succeq 0$ means $X$ positive semidefinite. This problem is convex and is also known as a semidefinite feasibility problem. Reasonably large instances can be solved efficiently using the algorithms detailed in \cite{Nest94} or \cite{Boyd03} for example.

\section{No Arbitrage Conditions}
\label{s:no-arbitrage}
In this section, we begin with an introduction on harmonic analysis on semigroups, which generalizes the moment conditions of the previous section to arbitrary payoffs. We then state our main result on the equivalence between no arbitrage in the one period market and positivity of the price matrices for the products in the payoff semigroup $\mathbb{S}$ defined in (\ref{eq:semigroup}):

\[
\mathbb{S}=\left\{1,x_1,\ldots,x_n,s_1(x),\ldots,s_m(x),x_1^2,\ldots,x_is_j(x),\ldots,s_m(x)^2,\ldots\right\}.
\]

\subsection{Harmonic analysis on semigroups}
We start by a brief primer on harmonic analysis on
semigroups (based on \cite{Berg84b} and the
references therein). Unless otherwise specified, all measures are
supposed to be positive.

A function $\rho(s):\mathbb{S}\rightarrow\reals$ on a semigroup $(\mathbb{S},\cdot)$ is called a \textit{semicharacter} if and only if it satisfies $\rho(st)=\rho(s)\rho(t)$ for all $s,t \in \mathbb{S}$ and $\rho(1)=1$. The dual of a semigroup $\mathbb{S}$, i.e. the set of semicharacters on $\mathbb{S}$, is written $\mathbb{S}^*$. 
\begin{definition} \emph{A function $f(s):\mathbb{S}\rightarrow\reals$ is a \textit{moment function on $\mathbb{S}$} if and only if $f(1)=1$ and $f(s)$ can be represented as:
\BEQ
f(s)=\int_{\mathbb{S}^*}{\rho(s)d \mu(\rho)},\quad\mbox{for all
}s\in\mathbb{S}, \label{eq:repres-moment}
\EEQ
where $\mu$ is a Radon measure on $\mathbb{S}^*$.} 
\end{definition}
When $\mathbb{S}$ is the semigroup defined in (\ref{eq:semigroup}) as an enlargement of the semigroup of monomials on $\reals^n$, its dual $\mathbb{S}^*$ is the set of applications $\rho_x(s):\mathbb{S}\rightarrow\reals$ such that $\rho_x(s)=s(x)$ for all $s\in\mathbb{S}$ and all $x\in\reals^n$. Hence when $\mathbb{S}$ is the payoff semigroup, to each point $x\in\reals^n$ corresponds a semicharacter that evaluates a payoff at that point. In this case, the condition $f(1)=1$ on the price of the cash means that the measure $\mu$ is a \emph{probability measure} on $\reals^n$ and the representation (\ref{eq:repres-moment}) becomes:
\BEQ
f(s)=\int_{\small\reals^n} s(x)d\mu(x)= \Expect_{\mu}\left[s(x)\right],\quad\mbox{for all
payoffs }s\in\mathbb{S}. \label{eq:repres-prob}
\EEQ
This means that when $\mathbb{S}$ is the semigroup defined in (\ref{eq:semigroup}) and there is no arbitrage, a moment function is a function that for each payoff $s\in\mathbb{S}$ returns its \emph{price} $f(s)=\Expect_{\mu}\left[s(x)\right]$. Testing for no arbitrage is then equivalent to testing for the existence of a moment function $f$ on $\mathbb{S}$ that matches the market prices in (\ref{eq:exist-proba}).

\begin{definition} \emph{A function $f(s):\mathbb{S} \rightarrow \reals$ is called \textit{positive semidefinite} if and only if for all finite families $\{s_i\}$ of elements of $\mathbb{S}$, the matrix with coefficients $f(s_i s_j)$ is positive semidefinite.} 
\end{definition}
We remark that moment functions are necessarily positive semidefinite. Here, based on results by \cite{Berg84b}, we  exploit this property to derive necessary and sufficient conditions for representation (\ref{eq:repres-prob}) to hold.

The central result in \cite[Th. 2.6]{Berg84b} states that the set of exponentially bounded positive semidefinite functions $f(s):\mathbb{S}\rightarrow \reals$ such that $f(1)=1$ is a Bauer simplex whose extreme points are given by the semicharacters in $\mathbb{S}^*$. Hence a function $f$ is positive semidefinite and exponentially bounded if and only if it can be represented as $f(s)=\int_{\mathbb{S}^*}{\rho d\mu(\rho)}$ with the support of
$\mu$ included in some compact subset of $\mathbb{S}^*$. 
Bochner' theorem on the Fourier transform of positive measures and Berstein's corresponding theorem for the Laplace transform are particular cases of this representation result. In what follows, we use it to derive tractable necessary and sufficient conditions for the function $f(s)$ to be represented as in (\ref{eq:repres-prob}).

\subsection{Main Result: No Arbitrage Conditions} 
We assume that the asset payoffs are bounded and that $\mathbb{S}$ is the payoff semigroup defined in (\ref{eq:semigroup}), this means that without loss of generality, we can assume that the payoffs $s_j(x)$ are positive. To simplify notations here, we define the functions $e_i(x)$ for $i=1,\ldots,m+n$ and $x\in\reals^n_+$ such that $e_i(x)=x_i$ for $i=1,\ldots,n$ and $e_{n+j}(x)=s_j(x)$ for $j=1,\ldots,m$. 

\begin{theorem} 
\label{th:no-arbitrage}
There is no arbitrage in the one period market and there exists a state price measure $\mu$ such that:
\[
\BA{l}
\Expect_\mu[x_i]=p_i,\quad i=1,\ldots,n,\\
\Expect_\mu[s_j(x)]=p_{n+j},\quad j=1,\ldots,m,\\
\EA
\]
if and only if there exists a function $f(s):\mathbb{S} \rightarrow \reals$ satisfying:
\begin{enumerate}
\item[(i)] $f(s)$ is a positive semidefinite function of $s\in\mathbb{S}$,
\item[(ii)] $f(e_is)$ is a positive semidefinite function of $s\in\mathbb{S}$ for $i=1,\ldots,n+m,$
\item[(iii)] $\left(\beta f(s) - \sum_{i=1}^{n+m}f(e_is)\right)$ is a positive semidefinite function of $s\in\mathbb{S}$,
\item[(iv)] $f(1)=1$ and $f(e_{i})=p_i$ for $i=1,\ldots,n+m,$
\end{enumerate}
for some (large) constant $\beta>0$, in which case we have $f(s)=\textstyle \Expect_{\mu}[s(x)]$.
\end{theorem}
\begin{proof}
By scaling $e_i(x)$ we can assume without loss of generality that $\beta=1$. For $s,u$ in $\mathbb{S}$, we note $E_s$ the shift operator such that for $f(s):\mathbb{S} \rightarrow \reals$, we have $E_u(f(s))\triangleq f(su)$ and we let $\mathcal{E}$ be the commutative algebra generated by the shift operators on $\mathbb{S}$. The family of shift operators $\tau=\{\{E_{e_i}\}_{i=1,\ldots,n+m},\left( I - \sum_{i=1}^{n+m}{E_{e_i}}\right)\}\subset\mathcal{E}$ is such that $I-T\in\mathrm{span}^+\tau$ for each $T\in\tau$ and
$\mathrm{span}\;\tau=\mathcal{E}$, hence $\tau$ is linearly 
admissible in the sense of \cite{Berg84a} or \cite{Mase77}, which states that (ii) and (iii) are equivalent to $f$ being $\tau$-positive. Then, \cite[Th. 2.1]{Mase77} means that $f$ is $\tau$-positive if and only if there is a measure $\mu$ such that $f(s)=\int_{\mathbb{S}^*}{\rho(s)d \mu(\rho)}$, whose support is a compact subset of the $\tau$-positive semicharacters. This means in particular that for a semicharacter $\rho_x \in \mathrm{supp}(\mu)$ we must have $\rho_x(e_i)\geq 0$, for $i=1,\ldots,n$ hence $x\geq0$. If $\rho_x$ is a $\tau$-positive
semicharacter then we must have $\{x\geq0:\|x\|_1\leq 1\}$, hence $f$ being $\tau$-positive is equivalent to $f$ admitting a representation of the form $f(s)=\Expect_{\mu}\left[s(x)\right]$, for all $s\in\mathbb{S}$ with $\mu$ having a compact support in a subset of the unit simplex.
\end{proof}

\subsection{Market Completeness}
As we will see below, under technical conditions on the asset prices, the moment problem is determinate and there is a one-to-one correspondence between the price $f(s)$ of the assets in $s\in\mathbb{S}$ and the state price measures $\mu$, in other words, the payoffs in $\mathbb{S}$ make the market \emph{complete}. 

Here, we suppose that there is no arbitrage in the one period market. Theorem \ref{th:no-arbitrage} shows that there is at least one measure $\mu$ such that $f(s)= \Expect_{\mu}\left[s(x)\right]$, for all payoffs $s\in\mathbb{S}$. In fact, we show below that when asset payoffs have compact support, this pricing measure is unique.

\begin{theorem}
\label{th:completeness}
Suppose that the asset prices $x_i$ for $i=1,\ldots,n$ have compact support, then for each set of arbitrage free prices $f(s)$ there is a unique state price measure $\mu$ with compact support satisfying:
\[
f(s)= \Expect_{\mu}\left[s(x)\right],\quad\mbox{for all
payoffs }s\in\mathbb{S}.
\]
\end{theorem}
\begin{proof}
If there is no arbitrage and asset prices $x_i$ for $i=1,\ldots,n$ have compact support, then the prices $f(s)= \Expect_{\mu}\left[s(x)\right]$, for $s\in\mathbb{S}$ are exponentially bounded in the sense of \cite[\S4.1.11]{Berg84b} and \cite[Th. 6.1.5]{Berg84b} shows that the measure $\mu$ associated to the market prices $f(s)$ is unique.
\end{proof}

This result shows that the securities in $\mathbb{S}$ make the market complete in the compact case. 

\subsection{Implementation}
\label{ss:implementation}
The conditions in theorem \ref{th:no-arbitrage} involve testing the positivity of infinitely large matrices and are of course not directly implementable. In practice, we can get a reduce set of conditions by only considering elements of $\mathbb{S}$ up to a certain (even) degree $2d$:
\BEQ
\label{eq:semigroup-deg}
\mathbb{S}_d\triangleq\left\{1,x_1,\ldots,x_n,s_1(x),\ldots,s_m(x),x_1^2,\ldots,x_is_j(x),\ldots,s_m(x)^2,\ldots,s_m(x)^{2d}\right\}
\EEQ
We look for a moment function $f$ satisfying conditions (i) through (iv) in Theorem \ref{th:no-arbitrage} for all elements $s$ in the reduced semigroup $\mathbb{S}_d$. Conditions (i)-(iii) now amount to testing the positivity of matrices of size $N_d={n+m+2d\choose n+m}$ or less. Condition (i) for example is written:
\[
\arraycolsep 1pt
\left(\BA{cccccccc} 
1 & p_1 & \cdots & p_{m+n} & f\left(x_1^2\right) & \cdots & f\left(s_m(x)^{\frac{N_d}{2}}\right)\\
p_1 & f\left(x_1^2\right) & \cdots & f\left(x_1s_m(x)\right) & f\left(x_1^3\right)& \cdots & f\left(x_1s_m(x)^{\frac{N_d}{2}}\right)\\
\vdots & \vdots & \ddots\\
p_{m+n} & f\left(x_1s_m(x)\right)&&&\vdots\\
f\left(x_1^2\right)& f\left(x_1^3\right)& & \cdots & f\left(x_1^4\right)\\
\vdots&\vdots&&&&&\vdots\\
f\left(s_m(x)^{\frac{N_d}{2}}\right) & f\left(x_1s_m(x)^{\frac{N_d}{2}}\right)&&&  & \cdots & f\left(s_m(x)^{N_d}\right)\\
\EA\right)\succeq 0,
\]
because the market price conditions in (\ref{eq:exist-proba}) impose $f(x_i)=p_i$ for $i=1,\ldots,n$ and  $f(s_j(x))=p_{n+j}$ for $j=1,\ldots,m$. Condition (ii) stating that $f(x_1s)$ be a positive semidefinite function of $s$ is then written as:
\[
\arraycolsep 2pt
\left(\BA{ccccc}
p_1 & f\left(x_1^2\right) & f\left(x_1x_2\right) & \cdots & f\left(x_1s_m(x)^{\frac{N_d}{2}-1}\right)\\
f\left(x_1^2\right) & f\left(x_1^4\right) & f\left(x_1^3x_2\right) & \\  
f\left(x_1x_2\right)& f\left(x_1^3x_2\right) & f\left(x_1^2x_2^2\right) \\ 
\vdots & & & \ddots & \vdots\\ 
f\left(x_1s_m(x)^{\frac{N_d}{2}-1}\right) & & & \cdots & f\left(x_1^2s_m(x)^{N_d-2}\right)\\
\EA\right)\succeq 0,
\]
and the remaining linear matrix inequalities in conditions (ii) and (iii) are handled in a similar way. These conditions are a finite subset of the full conditions in theorem \ref{th:no-arbitrage} and form a set of linear matrix inequalities in the values of $f(s)$ (see $\S\ref{ss:sdp}$). The exponential growth of $N_d$ with $n$ and $m$ means that only small problem instances can be solved using current numerical software. This is partly because most interior point based semidefinite programming solvers are designed for small or medium scale problems with high precision requirements. Here instead, we need to solve large problems which don't require many digits of precision.

\subsection{Multi-Period Models}
Suppose now that the products have multiple maturities $T_1,\ldots,T_q$. We know from \cite{Harr79} and \cite{Harr81} that the absence of arbitrage in this dynamic market is equivalent to the existence of a martingale measure on the assets $x_1,\ldots,x_n$. Theorem \ref{th:no-arbitrage} gives conditions for the existence of \emph{marginal} state price measures $\mu_i$ at each maturity $T_i$ and we need conditions guaranteeing the existence of a martingale measure whose marginals match these distributions $\mu_i$ at each maturity date $T_i$. A partial answer is given by the majorization result below, which can be traced to Blackwell, Stein, Sherman, Cartier, Meyer and Strassen. 
\begin{theorem}
If $\mu$ and $\nu$ are any two probability measures on a fininte set $A=\{a_1,\ldots,a_N\}$ in $\reals^N$ such that $\Expect_\mu[\phi]\geq\Expect_\nu[\phi]$ for every continuous concave function $\phi$ defined on the convex hull of $A$, then there is a martingale transition matrix $Q$ such that $\mu Q=\nu$.
\end{theorem}
Finding \emph{tractable} conditions for the existence of a martingale measure with given marginals, outside of the particular case of European call options considered in \cite{Davi05} for example, remains an open problem.

\section{Example: Spread Options}
To illustrate the results of section \ref{s:no-arbitrage}, we explicitly treat the case of a one period market with two assets $x_1$, $x_2$ with positive, bounded payoff at maturity and price $p_1,p_2$ today. European call options with payoff $(x-K_i)^+$ for $i=1,2$, are also traded on each asset with prices $p_3$ and $p_4$. We are interested in computing bounds on the price of a spread option with payoff $(x_1-x_2-K)^+$ given the prices of the forwards and calls.

We first notice that the complexity of the problem can be reduced by considering straddle options with payoffs $|x_i-K_i|$ instead of calls. Because a straddle can be expressed as a combination of calls, forwards and cash:
\[
|x_i-K_i|=(K_i-x_i)+2(x_i-K_i)^+.
\]
The advantage of using straddles is that the square of a straddle is a polynomial in the payoffs $x_i$, $i=1,2$. Using straddles instead of calls very significantly reduces the number of elements in the semigroup $\mathbb{S}_d$: when $k$ option prices are given on $2$ assets, this number is $(k+1){2+2d \choose 2}$, instead of ${2+k+2d \choose n+k}$. The payoff semigroup $\mathbb{S}_d$ is now:
\[
\mathbb{S}_d = \left\{ 1, x_1,x_2,|x_1-K_1|, |x_2-K_2|,|x_1-x_2-K|,\ldots,x_1|x_1-K_1|,\ldots,x_2^{2d}\right\}
\]
By sampling the conditions in theorem \ref{th:no-arbitrage} on $\mathbb{S}_d$ as in section \ref{ss:implementation}, we can compute a lower bound on the minimum (resp. an upper bound on the maximum) price for the spread option compatible with the absence of arbitrage. This means that we get an upper bound on the solution of:
\[\BA{ll}
\mbox{maximize} & \Expect_\mu[|x_1-x_2-K|]\\
\mbox{subject to} & \Expect_\mu[|x_i-K_i|]=p_{i+2}\\
& \Expect_\mu[x_i]=p_{i},\quad i=1,2\\
\EA\]
by solving the following program:
\[
\BA{lll}
\mbox{maximize} & f(|x_1-x_2-K|)\\
{ }\\
\mbox{subject to} & 
\left(\BA{cccc}
1&p_1&\cdots & f\left(x_2^d\right)\\
p_1&f\left(x_1^2\right)\\
\vdots & & \ddots & \vdots\\
f\left(x_2^d\right)& & \cdots & f\left(x_2^{2d}\right)\\
\EA\right)& \succeq0\\
& &\vdots\\
&\left(\BA{cccc}
f(b(x))&f(b(x)x_1)&\cdots & f\left(b(x)x_2^{d-1}\right)\\
f(b(x)x_1)&f\left(b(x)^2x_1^2\right)\\
\vdots & & \ddots & \vdots\\
f\left(b(x)x_2^{d-1}\right)& & \cdots & f\left(b(x)^2x_2^{2(d-1)}\right)\\
\EA\right)&\succeq0
\\
\EA
\]
where
\[
b(x)=\beta - x_1 - x_2 -|x_1-K_1| - |x_2-K_2| - |x_1-x_2-K|,
\]
is coming from condition (iii) in Theorem \ref{th:no-arbitrage}. This is a semidefinite program (see $\S\ref{ss:sdp}$) in the values of $f(s)$ for $s\in\mathbb{S}_d$.

\section{Conclusion}
By interpreting the \cite{Arro54} no arbitrage conditions as a moment problem, we have derived equivalent conditions directly written on the price of tradeable assets instead of state prices. This also shows how allowing trading in the products of market payoffs completes the market.




\bibliographystyle{agsm}
\bibliography{MainPerso}

\end{document}

%% file: defs.tex
\newcommand{\BEAS}{\begin{eqnarray*}}
\newcommand{\EEAS}{\end{eqnarray*}}
\newcommand{\BEA}{\begin{eqnarray}}
\newcommand{\EEA}{\end{eqnarray}}
\newcommand{\BEQ}{\begin{equation}}
\newcommand{\EEQ}{\end{equation}}
\newcommand{\BIT}{\begin{itemize}}
\newcommand{\EIT}{\end{itemize}}
\newcommand{\BNUM}{\begin{enumerate}}
\newcommand{\ENUM}{\end{enumerate}}

\newcommand{\BA}{\begin{array}}
\newcommand{\EA}{\end{array}}

\newcommand{\reals}{{\mathbb{R}}}

\newcommand{\Expect}{\textstyle\mathop{\bf E{}}}

\newcommand{\QED}{~~\rule[-1pt]{6pt}{6pt}}

\newtheorem{theorem}{Theorem}

\newtheorem{remark}[theorem]{Remark}

\newtheorem{definition}[theorem]{Definition}

\newcounter{exno}

\newenvironment{proof}{\textbf{Proof.}}{\QED\bigskip}

{\begin{quote}}{\end{quote}}

\makeatletter
\long\def\@makecaption#1#2{
   \vskip 9pt 
   \begin{small}
   \setbox\@tempboxa\hbox{{\bf #1:} #2}
   \ifdim \wd\@tempboxa > 5.5in
        \begin{center}
        \begin{minipage}[t]{5.5in}
        \addtolength{\baselineskip}{-0.95pt}
        {\bf #1:} #2 \par
        \addtolength{\baselineskip}{0.95pt}
        \end{minipage}
        \end{center}
   \else 
    \hbox to\hsize{\hfil\box\@tempboxa\hfil}  
   \fi
   \end{small}\par
}
\makeatother

\newcounter{oursection}

\newcounter{lecture}